

\magnification=1200
\baselineskip=20pt
\centerline{\bf Superconformal Covariantization Of Superdifferential Operator}
\centerline{\bf On $(1|1)$ Superspace And Classical N=2 W-superalgebras}
\vskip 1.5cm
\centerline{Wen-Jui Huang}
\vskip 1 cm
\centerline{Department of Physics}
\centerline{National Tsing Hua University}
\centerline{Hsinchu, Taiwan R.O.C.}
\vskip 1.5cm
\centerline{\bf Abstract}
\vskip 1cm

A study of the superconformal covariantization of superdifferential operators
defined
on $(1|1)$ superspace is presented. It is shown that a superdifferential
operator with a particular type of constraint can be covariantized only when
it is of
odd order. In such a case, the action of superconformal transformation on the
superdifferential operator is nothing but a hamiltonian flow
defined by the corresponding supersymmetric second
Gelfand-Dickey bracket. The covariant form of a superdifferential operator
of odd order is given.
\vskip 1cm
\centerline{\bf To appear in J. Math. Physics}
\vfil\eject
\noindent{\bf 1. Introduction}

Since Zamolochikov introduced the W-algebras[1], W-algebras
and related topics attracted a lot of
attention[2-12]. Not long after Zamolochikov's work  it was realized that
the classical versions of these algebras arise naturally in the
context of integrable systems in 1+1 dimension[2,7,8].
Indeed, the second hamiltonian structure of the nth order $KdV$
hierarchy provides a classical version of $W_n$-algebra.
In the Lax formulation,
the second hamiltonian structure is expressed elegantly by the so-called
second Gelfand-Dickey bracket associated with the corresponding
differential operator[13-16]. Recently, it was shown that the second
Gelfand-Dickey bracket associated with a pseudodifferential operator
also defines a hamiltonian structure and that the KP hierarchy is
hamiltonian with respect to it[17-22]. Here, we have a different class of
W-type algebras (called $W_{KP}$ algebras) from the second
Gelfand-Dickey bracket. More recently, the supersymmetric version
of the Gelfand-Dickey brackets have been constructed[23-25]. It was
discovered that the supersymmetric second Gelfand-Dickey bracket
associated with an odd-order superdifferential operator
on $(1|1)$ superspace gives (upon reduction) a superalgebra
which contain the classical N=2 super Virasoro algebra as
a subsuperalgebra. The analysis of the spectrum the
simplest case suggests that the resulted superalgebras
are N=2 W superalgebras[24,25]. However, a rigorous proof of
this statement is lacking. It is the purpose of this paper to
 set up a formalism which could help us analyze
the content of these superalgebras. More precisely, we shall
study the possibility of covariantizing the superdifferential
operators defined on $(1|1)$ superspace.

To see why the covariantization of superdifferential operator
is related to the spectra of the algebras resulting from
the corresponding supersymmetric Gelfand-Dickey bracket,
let us recall what we have learned in its bosonic counterpart.
We know that the definition of the W-type algebra requires
that it must contain a Virasoro
subalgebra and a set of primary fields of spin higher  than 2. For
instance, the $W_n$ algebra has, besides a Virasoro generator,
primary fields of spin up to $n$. On the other hand, each $W_{KP}$-type
algebra has a Virasoro generator and  primary fields of spin
up to $\infty$. However, the Gelfand-Dickey brackets are expressed in
terms of coefficient functions of the correponding (pseudo)differential
operators, which are generally not primary fields. One therefore has
to examine whether or not the required primary fields can be
constructed as	differential polynomials of these coefficient
functions. This task has been done in refs[7-10,26].
The proofs rely on the possibility of covariantizing the
corresponding (pseudo)differential operator. When a  (pseudo)differential
operator is properly covariantized the decompositions of the
coefficient functions into primary fields then follow immediately.
This suggests that the superconformal covariantization of
superdifferential operator could be helpful for analyzing the
spectra of the superalgebras from the supersymmetric Gelfand-Dickey
brackets. Unfortunately, as we shall see later, this program
does not completely solve the spectrum problem. The reason for it
is the fact that we are dealing with $N=2$ superalgebras while
our differential operators are defined on $(1|1)$ superspace and
the coefficient functions are $N=1$ superfields. As a result,
there seems no natural way to identify the needed $N=2$
supermultiplets. To be explicit, even though the flow
generated by the super Virasoro generator (which is $N=1$ superfield)
allows a geometrical interpretation on $(1|1)$ superspace
the flow generated by its superpartner, the superconformal primary
field of spin 1, does not. It is therefore necessary to find
a different approach to handle the effect of this spin-1 flow  in a
systematical way.

We organize this paper as follow. In Sec. 2 we describe the supersymmetric
second Gelfand-Dickey bracket briefly and derive the needed formulae.
In Sec. 3 we show that an appropriate covariance condition can be imposed
on a superdifferential operator of odd order and that the
resulted flow is nothing but the super Virasoro flow defined by the
corresponding supersymmetric Gelfand-Dickey bracket.  In Sec. 4  we
construct a sequence of covariant operators which are then used to
decompose the coefficient functions into primary fields. In Sec. 5  we
apply the result of Sec. 4 to study the simplest case in some details.
Finally, we offer some concluding remarks in Sec. 6.
\vskip 1cm

\noindent{\bf 2. Supersymmetric Gelfand-Dickey Bracket}

In this section we review briefly the supersymmetric second Gelfand-Dickey
bracket for later uses. We follow the conventions used in ref.[23]. We will
consider the superdifferential operators on a $(1|1)$ superspace with
coordinate $(x, \theta)$.  These operators are polynomials in the
supercovariant derivative $D=\partial_{\theta} + \theta \partial_x$ whose
coefficients are $N=1$ superfields;i.e.
$$ L= D^n + U_1 D^{n-1} + U_2 D^{n-2} + \dots +U_n  \eqno(2.1)$$
These operators are assumed to	be homogeneous under the usual $Z_2$ grading;
that is, $|U_i|=i$(mod 2). The bracket will involve functional of the form
$$ F[U] = \int_B f(U) \eqno(2.2)$$
where $f(U)$ is a homogeneous (under $Z_2$ grading) differential polynomial of
the $U_i$'s and $\int_B = \int dx d\theta$ is the Berezin integral which is
defined in the usual way, namely, if we write $U_i=u_i + \theta v_i$ and
$f(U)=a(u,v)+\theta b(u,v)$ then $\int_B f(U) = \int dx b(u,v)$. The
multiplication is given by the super Leibnitz rule:
$$ D^k \Phi = \sum^{\infty}_{i=0} \left[ \matrix{k \cr i} \right]
(-1)^{|\Phi|(k-i)} \Phi^{[i]} D^{k-i}, \eqno(2.3)$$
where $k$ is an arbitrary integer and $\Phi^{[i]} = (D^i \Phi)$
and the superbinomial coefficients $\left[ \matrix{k \cr i} \right]$ are
defined by
$$\left[ \matrix{k \cr i} \right] = \left\{ \eqalign{ & 0 \qquad for \quad i>k
\quad or \quad (k,i) \equiv (0,1) \quad (mod \quad 2) \cr
& \pmatrix{ [{k \over 2}] \cr [{i \over 2}] }  \qquad otherwise \cr}
\right\} \eqno(2.4)$$
where $\pmatrix{p \cr q}$ is the ordinary binomial coefficient.
Next, we introduce the notions of superresidue and supertrace. Given a
super-pseudodifferential operator $P=\sum p_i D^i$ we define its superresidue
as $$ sresP=p_{-1} \eqno(2.5)$$
and its supertrace as
$$ StrP = \int_B sresP. \eqno(2.6)$$
In the usual manner it can be shown that the supertrace of a supercommutator
vanishes;i.e.
$$ Str[P,Q] = 0 \eqno(2.7)$$
where
$$ [P,Q] \equiv PQ-(-1)^{|P||Q|}QP. \eqno(2.8)$$
Finally, for a given functional $F[U]=\int_B f(U)$ we define its gradient $dF$
by
$$ dF=\sum^n_{i=1} (-1)^{n+k} D^{-n+k-1} {\delta f \over \delta U_k},
\eqno(2.9)$$
where
$$ {\delta f \over \delta U_k} = \sum^{\infty}_{i=0} (-1)^{|U_k|i+i(i+1)/2}
D^i {\partial f \over \partial U_k^{[i]}}. \eqno(2.10)$$
Equipped with these notions we now define the supersymmetric second
Gelfand-Dickey bracket as
$$\{F,G\}=(-1)^{|F|+|G|+n} Str[L(dFL)_+dG - (LdF)_+LdG] \eqno(2.11)$$
where $( )_+$ denotes the differential part of a supe-pseudodifferential
operator. It has been shown that (2.11) indeed defines a hamiltonian structure:
it is antisupersymmetric and satisfies the super-Jacobi identity[23].

In ref.[24] it is shown that when the constraint $U_1=0$ is imposed the
induced bracket is well-defined only when $n$ is odd. The reason is that this
constraint is second class when $n$ is odd, while becomes first class for even
$n$'s. To describe these induced brackets, we need to modify at least one of
$dF$ and $dG$ defined by (2.9) due to absence of $U_1$. The prescription is
to add a term $D^{-n} V$ to, say, dG in such a way that
$$ sres[L,D^{-n} V + dG] =0 \eqno(2.12)$$
We shall denote $X_G = D^{-n} V + dG$ for this choice of $V$. Replacing $dG$
in (2.11) by $X_G$ then gives the induced bracket. A useful operator form of
the induced bracket is
$$\eqalign{ J(X_G) &= (LX_G)_+L - L(X_GL)_+ \cr
		   &= \sum^n_{i=2} (-1)^{k|G|+1} \{U_k,G\} D^{n-k}
\cr}\eqno(2.13)$$
We shall also regard $J(X_G)$ as the transformation of the superdifferential
operator $L$ under the hamiltonian flow defined by $G$.

It is known that if we define
$$\eqalign{ T &= U_3 - {1 \over 2} U'_2 \cr
	    J &= U_2 \cr}\eqno(2.14)$$
where $V'=(DV), V''=(D^2V), \dots$ etc, then $T$ and $J$ obey[24]
$$\eqalign{ \{T(X), T(Y)\} &= [{1 \over 4}m(m+1)D^5 + {3 \over 2} T(X) D^2 +
{1 \over 2}T'(X)D + T''(X)] \delta(X-Y), \cr
\{T(X),J(Y)\} &= [-J(X)D^2 + {1 \over 2}J'(X)D -{1 \over 2}J''(X)] \delta(X-Y),
\cr \{J(X),T(Y)\} &= [J(X)D^2 -{1 \over 2}J'(X)D + J''(X)] \delta(X-Y), \cr
\{J(X),J(Y)\} &= -[m(m+1)D^3 + 2T(X)] \delta(X-Y), \cr} \eqno(2.15)$$
where we have written $n=2m+1$ and $\delta(X-Y)=\delta(x-y)(\theta-w)$.
(2.15) is
the classical $N=2$ super Virasoro algebra. It is conjectured that each
remaining field $U_j$ for $j$ even gives rise to an $N=2$ superconformal
primary field $W_j$ abtained by deforming $U_j$ via the addition of
differential  polynomials in the $U_{i<j}$ and that the remaining $U_j$ with
$j$ odd give rise to their partners. This conjecture naturally leads us to
consider the hamiltonian flows defined by the two linear functionals:
$$\eqalign{ G &= \int_B T\xi = \int_B (U_3 \xi + {1 \over 2} U_2 \xi') \cr
	    H &= \int_B J\zeta = \int_B U_2 \zeta \cr} \eqno(2.16)$$
where $|\xi(x,\theta)|=|\zeta(x,\theta)|=0$.
Putting (2.16) into (2.13) we obtain
$$\eqalign{J(X_G) &= [\xi D^2 + {1 \over 2}\xi'D + {(m+1) \over 2}\xi'']L -
		     L[\xi D^2 + {1 \over 2}\xi'D -{m \over 2}\xi''] \cr
J(X_H) &= [-\zeta D-(m+1)\zeta']L - L[-\zeta D + m \zeta'] \cr}\eqno(2.17)$$
Since $T$ is the super Virasoro generator, $J(X_G)$ is called the super
Virasoro flow.
We shall prove in the next section that $J(X_G)$ in (2.17) arises quite
naturally once we impose on $L$ a covariance condition which amounts to
requiring $L$ to satisfy a particular transformation law under the
superconformal transformation on $(1|1)$ superspace.
\vskip 1cm

\noindent{\bf 3. Superconformal Covariance And Super Virasoro Flow}

Let us consider the $(1|1)$ superspace with coordinate $X=(x,\theta)$.
The most general superdiffeomorphism has the form
$$\eqalign{\tilde{x} &= g(x) + \theta \kappa(x) \cr
\tilde{\theta} &= \chi(x) + \theta B(x) \cr} \eqno(3.1)$$
where $|g|=|B|=0$ and $|\kappa|=|\chi|=1$.
Under the superdiffeomorphism (3.1) the superderivative transforms as follows:
$$ D = (D \tilde{\theta}) \tilde{D}+ [(D \tilde{x})-\tilde{\theta} (D
\tilde{\theta})]
(\tilde{D})^2 \eqno(3.2)$$
We call the superdiffeomrphism (3.1) a superconformal transformation if
$$ D = (D \tilde{\theta}) \tilde{D} \eqno(3.3)$$
or, equivalently,
$$ D \tilde{x} = \tilde{\theta} (D \tilde{\theta}) \eqno(3.4)$$
A function $f(X)$ is called a superconformal primary field of spin $h$ if,
under superconformal transformation, it transforms as
$$ f(\tilde{X}) = (D \tilde{\theta})^{-2h} f(X) \eqno(3.5)$$
We shall denote by $F_h$ the space of all superconformal primary fields of
spin $h$. As usual, a superdifferential operator $\Delta$ is called  a
covariant operator if it maps $F_h$ to $F_l$ for some $h$ and $l$.

We are ready to study the covariant property of the superdifferential operator
$$ L = D^n + U_2 D^{n-2} + U_3 D^{n-3} + \dots + U_n \eqno(3.6)$$
where we have set $U_1$ to be zero. As in the bosonic case, we like to impose
the covariance condition:
$$ L \quad : \quad  F_h \longrightarrow F_l \eqno(3.7)$$
or, equivalently,
$$ L(X') = (D \tilde{\theta})^{-2l} L(X) (D \tilde{\theta})^{2h}  \eqno(3.8)$$
for some $h$ and $l$. In other words, we like to see if there exists a
transformation of the functions $U_2, \dots, U_n$ such that the operator $L$
is a covariant operator. We expect, as in the bosonic case, the constraint
$U_1=0$ determines both the values $h$ and $l$. To this purpose, we rewrite
(3.8) as
$$ L(X') (D \tilde{\theta})^{-2h} = (D \tilde{\theta})^{-2l} L(X)  \eqno(3.9)$$
By using (3.3) the first term on the left hand side of (3.9) can be expanded as
$$ (\tilde{D})^n (D \tilde{\theta})^{-2h} = (D \tilde{\theta})^{-2h-n} (D^n +
A_{n-1} D^{n-1} + A_{n-2} D^{n-2} + \dots) \eqno(3.10)$$
With simple algebras we find
$$ A_{n-1} = \left\{ \eqalign{& m \qquad n=2m \cr
& -2h-m \qquad n=2m+1  \cr} \right\} \eqno(3.11)$$
Thus, for even $n$ the constraint $U_1=0$ cannot be preserved under
superconformal transformation. But when
$$ n=2m+1 \eqno(3.12)$$
the constraint is preserved if one chooses
$$ h= -{1 \over 4}(n-1) = -{1 \over 2} m  \eqno(3.13)$$
As a result of (3.9), (3.10)  and (3.13), the only choice of $l$ is then
$$ l= {1 \over 4}(n+1) = {1 \over 2}(m+1) \eqno(3.14)$$
With these choices the covariance condition (3.8) reads
$$ L(\tilde{X}) = (D \tilde{\theta})^{-(m+1)} L(X) (D \tilde{\theta})^{-m}
\eqno(3.15)$$ which then determines how the functions $U_k$'s transform under
superconformal transformation. For example,   simple computations yield
the transformation laws of $U_2$ and $U_3$:
$$\eqalign{U_2(X) &= U_2(\tilde{X}) (D \tilde{\theta})^2 \cr
U_3(X) &= U_3(\tilde{X}) (D \tilde{\theta})^3 + U_2(\tilde{X}) (D
\tilde{\theta}) (D^2 \tilde{\theta}) + {1 \over 2}m(m+1) S(\tilde{X},X) \cr}
\eqno(3.16)$$
where $S(\tilde{X},X)$ is the superschwarzian defined by
$$S(\tilde{X},X) ={{D^4 \tilde{\theta}} \over {D \tilde{\theta}}} -2 \big(
{{D^3
\tilde{\theta}} \over
{D \tilde{\theta}}} \big) \big({{D^2 \tilde{\theta}} \over {D \tilde{\theta}}}
\big) \eqno(3.17)$$
One recognizes at once that $U_2$ is a superconformal primary field of spin 1.
Moreover, using (3.16) we find that $T$ defined by (2.14) transforms as
$$ T(X) = T(\tilde{X}) (D \tilde{\theta})^3 + {1 \over 2}m(m+1) S(\tilde{X},X)
\eqno(3.18)$$ We therefore see that $T$ has the same transformation law as the
energy-momentum tensor in the superconformal theory. It is not hard to
verify that the
infinitesimal forms of (3.16) and (3.18) is the same as the corresponding
transformation laws from $J(X_G)$ of (2.17). As a matter of fact, we can
prove that, with a suitable identification of parameter $\xi$, the
infinitesimal form of (3.15) is precisely equal to $J(X_G)$.
To  prove this statement, we first write down the most general infinitesimal
form of superconformal transformation:
$$\eqalign{ \tilde{x} &= x- \epsilon(x) - \theta \eta(x) \cr
  \tilde{\theta} &= \theta - {1 \over 2} \partial_x \epsilon(x) \theta
- \eta(x) \cr} \eqno(3.19)$$
where $|\epsilon|=0$ and $|\eta|=1$. From now on we shall keep terms up to
linear in $\epsilon$ and $\eta$ in all computations. Define $\xi(x,\theta)=
{1 \over 2} \epsilon(x) + \theta \eta(x)$ we find
$$\eqalign{D \tilde{\theta} &= 1- \xi'' \cr
\tilde{D} &= D + \xi'' D = D + D[D, \xi]D + [D,\xi]D^2 \cr}\eqno(3.20)$$
By induction, we derive from (3.20) the formula:
$$(\tilde{D})^k = D^k + D[D^k,\xi]D + [D^k,\xi]D^2 \eqno(3.21)$$
which has a more useful equivalent form
$$ (\tilde{D})^k = D^k + D^k(\xi'D) -(\xi' D)D^k +2 [D^k,\xi]D^2 \eqno(3.22)$$
Secondly, we note
$$\eqalign{U_k(\tilde{X}) &= U_k(x-\epsilon - \theta \eta, \theta - {1 \over 2}
\partial_x \epsilon \theta - \eta) + \delta_{\xi} U_k \cr
 &= U_k(X) - 2 \xi \partial_x U_k - \xi' (DU_k) + \delta_{\xi} U_k \cr}
\eqno(3.23)$$
Now (3.20), (3.22) and (3.23) together yield
$$\eqalign{ (D \tilde{\theta})^{-m-1} L(X) (D \tilde{\theta})^{-m} &= (1-
\xi'')^{-m-1} L(X) (1- \xi'')^{-m} \cr
&= L(X) + (m+1) \xi'' L + m L \xi'' \cr} \eqno(3.24)$$
and
$$\eqalign{ L(\tilde{X}) &= L(X) - 2 \xi [\partial_x, L] -\xi'
\sum^{2m+1}_{k=2} (DU_k)D^{2m+1-k} + \delta_{\xi} L + L (\xi' D) \cr
& \qquad - \sum^{2m+1}_{k=2} U_k \xi' D D^{2m+1-k} + 2 [L, \xi] \partial_x \cr
&= L(X) - [2 \xi D^2 + \xi' D ] L + L [\xi' D + 2 \xi D^2] + \delta_{\xi} L
\cr} \eqno(3.25)$$
Equating (3.24) with (3.25) we obtain the infinitesimal form of (3.15):
$$\delta_{\xi} L = [2\xi D^2 + \xi' D + (m+1) \xi''] L -
 L [2 \xi D^2 + \xi' D -m \xi] \eqno(3.26)$$
which is equal to $J(X_G)$ given by (2.17), provided that the trivial
redefinition $\xi \longrightarrow {1 \over 2} \xi$ is taken. We therefore
have shown that the infinitesimal form of covariance condition is nothing but
the super Virasoro flow.
\vskip 1cm

\noindent{\bf 4. Superconformal Covariantization of $L$}

In this section we shall covariantize the superdifferential operator (3.6).
The construction will be parallel to that for the bosonic case. First, we
define
$$ B(\tilde{X},X) = {{D^2 \tilde{\theta}} \over {D \tilde{\theta}}}
\eqno(4.1)$$
We can show easily that $B(\tilde{X},X)$ has the following transformation law:
$$ B(\tilde{\tilde{X}},X) = (D \tilde{\theta}) B(\tilde{\tilde{X}},\tilde{X}) +
B(\tilde{X},X) \eqno(4.2)$$
and that the superschwarzian can
be represented as
$$ S(\tilde{X},X) = D^2 B(\tilde{X},X) - (D B(\tilde{X},X)) B(\tilde{X},X)
\eqno(4.3)$$
Using (4.2) we can verify that the superschwarzian satisfies
$$ S(\tilde{\tilde{X}},X) = (D \tilde{\theta})^3 S(\tilde{\tilde{X}},\tilde{X})
+ S(\tilde{X},X) \eqno(4.4)$$
Now we choose a particular coordinate $Z=(z, \vartheta)$  and demand
$$ T(X) = {m(m+1) \over 2} S(Z,X) \eqno(4.5)$$
The transformation law (4.4) then guarantees $T(X)$ transforms as (3.18).
Obviously, this choice of coordinate is to make $T$ vanish identically;i.e.
$T(Z)=0$. We are not going to concern with the problem of existence of	such
a coordinate, which is beyond the scope of this paper, but simply insist
the identification (4.5). For the rest of
this section we shall use the notation:
$$ B(X) \equiv B(Z,X) \eqno(4.6)$$
and the representation of $T$:
$$ T(X) = {m(m+1) \over 2} [D^2 B(X) - (D B(X)) B(X)] \eqno(4.7)$$
One should note that different $B(X)$'s may define the same $T(X)$. Indeed,
if we replace $B$ by $B + \delta B$ and demand $\delta B$ satisfy
$$ D^2 (\delta B) - [D(\delta B)]B - (DB) \delta B = 0 \eqno(4.8)$$
then $T$ is not changed.

The definition of $B(X)$ enables us to introduce a covariant superderivative
defined by
$$ \hat{D}_{2k} \equiv D - 2k B(X) \eqno(4.9)$$
One can verify easily that $\hat{D}_{2k}$ maps from $F_k$ to $F_{k+{1 \over
2}}$. Hence the operator
$$\eqalign{ \hat{D}^{l}_{2k} &\equiv \hat{D}_{2k+l-1} \hat{D}_{2k+l-2} \dots
\hat{D}_{2k}  \qquad (l>0) \cr
&=[D-(2k+l-1)B][D-(2k+l-2)B] \dots [D-2kB] \cr} \eqno(4.10)$$
maps from $F_k$ to $F_{k+{l \over 2}}$, that is, it transforms, under
superconformal transformation, as
$$\hat{D}^l_{2k} (\tilde{X}) = (D \tilde{\theta})^{-2k-l}
\hat{D}^l_{2k} (D \tilde{\theta})^{2k}
\eqno(4.11)$$
We list here  two useful relations following from the definition (4.9)
of covariant superderivative. The first one is
$$ \hat{D}_{2k} \delta B = - \delta B \hat{D}_{2k-1} + \triangle B
\eqno(4.12)$$
where  $\delta B$ is an {\it arbitrary} variation and
$$ \triangle B \equiv D(\delta B) - B \delta B	\eqno(4.13)$$
The other one  is an equivalent form of (4.8):
$$ \hat{D}_{2k+1} \hat{D}_{2k} \delta B = \delta B \hat{D}_{2k} \hat{D}_{2k-1}
\eqno(4.14)$$
By using (4.13) and (4.14) we can easily derive the variation of
$\hat{D}^l_{2k}$ due  to   $\delta B$ subjected to (4.8). The results are
$$ \delta_B \hat{D}^{2m}_{2k} = -\delta B (m \hat{D}^{2m-1}_{2k} ) -
 \triangle B [m(2k+m-1) \hat{D}^{2m-2}_{2k}] \eqno(4.15)$$
and
$$ \delta_B \hat{D}^{2m+1}_{2k} = - \delta B [(2k+m) \hat{D}^{2m}_{2k}]
 - \triangle B [m(2k+m) \hat{D}^{2m-1}_{2k} ] \eqno(4.16)$$
An important consequence of (4.15) and (4.16) is that the covariant operator
$\hat{D}^l_{2k}$ depends explicitly on $B$ except when $l=2m+1$ and $k=-{m
\over
2}$. In these exceptional cases, it depends on $B$ only through $T$. This
result,
of course, can be expected from (3.12) and (3.13). Now we are ready to
construct
covariant operators involving superconformal primary fields. Let us consider
$$ \Delta^{(2m+1)}_{2p} (W_{2p}, T) = \sum^{2m+1-2p}_{i=0} \alpha_{2p,i}
(\hat{D}^i_{2p} W_{2p}) \hat{D}^{2m+1-2p-i}_{-m}  \qquad \alpha_{2p,0}=1
\eqno(4.17)$$
where $W_{2p}$ is a superconformal primary field of spin $p$.
We like to choose $\alpha_{2p,i}$'s in such a way that the right hand side
of (4.16)
depends on $B$ only through $T$. To this end, we have to compute the variation
with respect to $B$ with $\delta B$ constrained by (4.14). For integer $p$
we find
$$ \delta_B \Delta^{(2m+1)}_{2p} \equiv  \delta B \big({{\delta
\Delta^{(2m+1)}_{2p}}
\over \delta B} \big) + \triangle B \big({{\delta \Delta^{(2m+1)}_{2p}} \over
\triangle B} \big) \eqno(4.18)$$
where
$$\eqalign{{{\delta \Delta^{(2m+1)}_{2p}} \over \delta B} = &-\sum^{m-p}_{i=1}
[i \alpha_{2p,2i} - (m-p-i+1) \alpha_{2p,2i-1} ] (\hat{D}^{2i-1}_{2p} W_{2p})
\hat{D}^{2(m-p-i)+1}_{-m}  \cr
&  \sum^{m-p+1}_{i=1} [(p+i-1)
\alpha_{2p,2i-2} - (2p+i-1) \alpha_{2p,2i-1}] (\hat{D}^{2i-2}_{2p} W_{2p})
\hat{D}^{2(m-p-i+1)}_{-m} \cr} \eqno(4.19)$$
and
$$\eqalign{&{{\delta \Delta^{(2m+1)}_{2p}} \over \triangle B} \cr
&= \sum^{m-p-1}_{i=0} [(p+i)(m-p-i) \alpha_{2p,2i} - (i+1)(2p+i)
\alpha_{2p,2i+2}]
(\hat{D}^{2i}_{2p} W_{2p} ) \hat{D}^{2(m-p-i)-1}_{-m} \cr
&\qquad - \sum^{m-p}_{i=1} [i(2p+i) \alpha_{2p,2i+1} -(m-p-i+1)(p+i)
\alpha_{2p,2i-1}] (\hat{D}^{2i-1}_{2p} W_{2p}) \hat{D}^{2(m-p-i)}_{-m} \cr}
\eqno(4.20)$$
Demanding ${{\delta \Delta^{(2m+1)}_{2p}} \over \delta B}=0$ and ${{\delta
\Delta^{(2m+1)}_{2p}} \over \triangle B}=0$ gives, respectively,
$$ \eqalign{ \alpha_{2p,2i+1} &= {{p+i} \over {2p+i}} \alpha_{2p,2i} \cr
     \alpha_{2p,2i} &= {(m-p-i+1) \over i} \alpha_{2p,2i-1} \cr}  \eqno(4.21)$$
and
$$ \eqalign{\alpha_{2p,2i+2} &= {(p+i)(m-p-i) \over (i+1)(2p+i)}
\alpha_{2p,2i} \cr
\alpha_{2p,2i+1} &= {(p+i)(m-p-i+1) \over i(2p+i)} \alpha_{2p,2i-1} \cr}
\eqno(4.22)$$
Remarkably, (4.21) implies (4.22). Therefore, $\alpha_{2p,i}$'s are determined
umabiguously. For a half integer $p=q+{1 \over 2}$ the calculation is much
the same. We simply write down the resulted recursion relateions:
$$ \eqalign{\alpha_{2q+1,2i} &= {{q+i} \over i} \alpha_{2q+1,2i-1} \cr
 \alpha_{2q+1,2i+1} &= {{m-q-i} \over {2q+i+1}}
\alpha_{2q+1,2i} \cr} \eqno(4.23)$$
Solving (4.21) and (4.23) then yields
$$ \eqalign{ \alpha_{2p,2l} &= (-1)^l {{\pmatrix{l+p-m-1 \cr l }
\pmatrix{p+l-1 \cr l }} \over \pmatrix{2p+l-1 \cr l } } \cr
\alpha_{2p,2l+1} &= {(-1)^l \over 2} {{\pmatrix{p+l-m-1 \cr l }
\pmatrix{p+l \cr l }} \over \pmatrix{2p+l \cr l } } \cr} \eqno(4.24)$$
and
$$ \eqalign{ \alpha_{2q+1,2l} &= (-1)^l {{\pmatrix{q+l-m-1 \cr l }
\pmatrix{q+l \cr l}} \over \pmatrix{2q+l \cr l }} \cr
\alpha_{2q+1,2l+1} &= (-1)^l {{m-q} \over 2q+1} {{\pmatrix{q+l-m \cr l }
\pmatrix{q+l \cr l }} \over \pmatrix{2q+l+1 \cr l }} \cr} \eqno(4.25)$$
With the coefficients $\alpha_{2p,l}$'s given by (4.24) and (4.25) we now
write
$$\eqalign{ L &= D^{2m+1} + U_2 D^{2m-1} + \dots  + U_{2m+1} \cr
&= \hat{D}^{2m+1}_{-m} + \Delta^{(2m+1)}_2(U_2, T) + \sum^{2m+1}_{k=4}
\Delta^{(2m+1)}_{k} (W_k, T) \cr} \eqno(4.26)$$
which is the desired covariant form. If one works out explicitly the right
hand side of (4.26), one would obtain decomposition of the form
$$ U_k = W_k + G_k(W_{k-1}, \dots, W_4, T, U_2)  \qquad (k \ge 4) \eqno(4.27)$$
where $G_k$ is a differential polynomial in $W_{k-1}, \dots, W_4, T, U_2$.
Inverting (4.27) gives the definitions of superconformal primary fields in
terms of coefficient functions:
$$ W_k = U_k + H_k(U_{k-1}, \dots, U_4, T, U_2)  \qquad (k \ge 4) \eqno(4.28)$$
where $H_k$ is again a differential polynomial. This completes the
covariantizaation of $L$.

Before ending this section we like to remark that so far we have only taken
care of super Virasoro flow. In other words, what we have done is to decompose
the coefficient functions into superconformal primary fields which satisfy
$$ \delta_{\xi} W_k = - \{ W_k, G\} = {k \over 2} W_k \xi'' +
{(-1)^{k+1} \over 2} W'_k
\xi' + W''_k \xi \eqno(4.29)$$
We do not  know yet how $W_k$'s transform under
$J(X_H)$, the flow generated by the spin-1 current $J$. As a result, we can
{\it not} expect that $W_{2k}$ and $W_{2k+1}$ do form a $N=2$ supermultiplet.
To identify the supermultiplets some redefinitions of primary fields should
be expected. For example, to obtain the first two supermultiplets the following
redefinitions should be considered
$$\eqalign{ \bar{W_4} &= W_4 + a J^2 \cr
	    \bar{W_5} &= W_5 \cr
	    \bar{W_6} &= W_6 + b J W_4 + c J^3 \cr
	    \bar{W_7} &= W_7 + e J W_5 \cr} \eqno(4.30)$$
where $a$, $b$, $c$ and $e$ are constants. Indeed, we shall see in the next
section, where the simplest nontrivial case is studied, that redefinitions
of this sort must be done in order to get  the desired supermultiplets.
\vskip 1cm

\noindent{\bf 5. An Explicit Example}

In this section we study the simplest nontrivial case:
$$ L = D^5 + U_2 D^3 + U_3 D^2 + U_4 D + U_5 \eqno(5.1)$$
Even though this case has been studied in the literature[25] we like to use
it to illustrate the usefulness of the results in the previous section.
By (4.26) we have
$$ L = \hat{D}^5_{-2} + \Delta^{(5)}_2(J,T) + \Delta^{(5)}_4(W_4,T) +
\Delta^{(5)}_5(W_5,T) \eqno(5.2)$$
By using (4.24) and (4.25) we find
$$\eqalign{ \hat{D}^5_{-2} &= (D-2B)(D-B)D(D+B)(D+2B) \cr
     &= D^5 + T D^2 + {1 \over 3}T' D + {2 \over 3} T'' \cr
\Delta^{(5)}_2 (J,T) &= J \hat{D}^3_{-2} + {1 \over 2}(\hat{D}_2 J)
\hat{D}^2_{-2} + {1 \over 2} (\hat{D}^2_2 J) \hat{D}_{-2} + {1 \over 3}
(\hat{D}^3_2   J) \cr
  &= J D^3 + {1 \over 2}J' D^2 + {1 \over 2} J'' D + {1 \over 3}J''' +
{4 \over 9} JT \cr
\Delta^{(5)}_4 (W_4,T) &= W_4 \hat{D}_{-2} + {1 \over 2} (\hat{D}_4 W_4) \cr
  &= W_4 D + {1 \over 2} W'_4 \cr
\Delta^{(5)}_5 (W_5,T) &= W_5 \cr} \eqno(5.3)$$
Thus, we have the following decompositions:
$$\eqalign{U_2 &= J \cr
U_3 &= T + {1 \over 2} J' \cr
U_4 &= W_4 + {1 \over 3}T' + {1 \over 2} J'' \cr
U_5 &= W_5 + {1 \over 2}W'_4 + {2 \over 3} T'' + {1 \over 3} J''' +
{4 \over 9} JT \cr} \eqno(5.4)$$
Inverting (5.4) then gives the definitions primary fields in terms of
coefficient functions:
$$\eqalign{J &= U_2 \cr
T &= U_3 - {1 \over 2}U'_2 \cr
W_4 &= U_4 - {1 \over 3} U'_3 - {1 \over 3} U''_2 \cr
W_5 &= U_5 - {1 \over 2} U'_4 - {1 \over 2} U''_3 + {1 \over 6} U'''_2 -
       {4 \over 9} TU_2 \cr} \eqno(5.5)$$
As we explained at the end of section 4 that $W_4$ and $W_5$ may not form a
$N=2$ supermultiplet. To check this point we have to compute $J(X_H)$ given by
(2.17) explicitly. Straightforward calculations yield
$$\eqalign{J(X_H) &= [- \zeta D - 3 \zeta'] L - L [- \zeta D + 2 \zeta'] \cr
 &\equiv (\delta_{\zeta} U_2) D^3 + (\delta_{\zeta} U_3) D^2 +
(\delta_{\zeta} U_4) D
 + (\delta_{\zeta} U_5) \cr} \eqno(5.6)$$
where
$$\eqalign{\delta_{\zeta} U_2 &= [ 6 D^3 + (2 U_3 - U'_2)] \zeta \cr
\delta_{\zeta} U_3 &= [- 3 D^4 - U_2 D^2 + U_3 D - U'_3 ] \zeta \cr
\delta_{\zeta} U_4 &= [3 D^5 + 3 U_2 D^3 + U_3 D^2 + (2 U_5 - U_4')] \zeta \cr
\delta_{\zeta} U_5 &= [-2 D^6 - 2 U_2 D^4 -2 U_3 D^3 - 2 U_4 D^2 + U_5 D -
U'_5] \zeta \cr} \eqno(5.7)$$
Using (5.4) and (5.5) we further find
$$\eqalign{\delta_{\zeta} J &= [6 D^3 + 2T] \zeta \cr
\delta_{\zeta} T &= [-J D^2 + {1 \over 2} J' D - {1 \over 2} J''] \zeta \cr
\delta_{\zeta} W_4 &= [{8 \over 3}J D^3 + {8 \over 9} JT + 2 W_5] \zeta \cr
\delta_{\zeta} W_5 &= [-2(W_4- {2 \over 9}J^2) D^2 + {1 \over 2} (W_4- {2 \over
9} J^2 )' D - {1 \over 2} (W_4-{2 \over 9}J^2)''] \zeta \cr} \eqno(5.8)$$
The first two equations of (5.8) give rise to, with the help of (2.13), Poisson
brackets as expected from (2.15). On the other hand, since $J$ and $T$ show up
in $\delta_{\zeta} W_4$ and $\delta_{\zeta} W_5$, $W_4$ and $W_5$ do not form
a $N=2$ supermultiplet. Hence, a redefinition of $W_4$ of the form of (4.30) is
necessary. Indeed, the last of (5.8) does suggest the following redefinition:
$$\eqalign{ \bar{W_4} &= W_4 - {2 \over 9} J^2 \cr
&= U_4 - {1 \over 3} U'_3 - {1 \over 3} U''_2 -{2 \over 9} U^2_2 \cr}
\eqno(5.9)$$
With (5.9) we then obtain
$$\eqalign{\bar{W_4} &= 2 W_5 \zeta \cr
W_5 &= [-2 \bar{W}_4 D^2 + {1 \over 2} \bar{W}'_4 D - {1 \over 2} \bar{W}''_4]
\zeta \cr} \eqno(5.10)$$
The corresponding Poisson brackets can be easily read off:
$$\eqalign{ \{\bar{W}_4(X), J(Y) \} &= -2 W_5 \delta(X-Y) \cr
\{ W_5(X), J(Y) \} &= [-2 \bar{W}_4 D^2 + {1 \over 2} \bar{W}'_4 D - {1 \over
2}  \bar{W}''_4 ] \delta(X-Y) \cr} \eqno(5.11)$$
We therefore conclude that $\bar{W}_4$ and $W_5$ form a $N=2$ supermultiplet.
\vskip 1cm

\noindent{\bf 6. Conclusions}

In this paper we have carried out the study of superconformal covariantization
of superdifferential operators. We have shown that when the constraint $U_1=0$
is imposed only those of odd order can be consistently covariantized. The
covariance condition is then shown to be equivalent to the super Virasoro flow.
As a result, the covariant form of a superdifferential operator immediately
leads to the decompositions of coefficient
functions into differential polynomials of
spin-1 supercurrent, super Virasoro generator and superconformal primary fields
of spin higher than ${3 \over 2}$. However, to prove the corresponding
superalgebra to be a $N=2$ W-superalgebra this is only half the way. The
essential point is that the superdifferential operators are defined on the
$(1|1)$ superspace and hence there is no natural way to interpret the flow
generated by the spin-1 supercurrent in a geometrical manner. As illustrated
by the simplest nontrivial example, explicit calculations and further
redefinitions of superconformal primary fields are required to identify the
desired $N=2$ supermultiplets. The problem of systemmatical identifications of
$N=2$ supermultiplets  for superdifferential operators of high orders therefore
remains open.

Finally we like to remark that there exists an interesting link between
the covariant differential operators and a class  of singular vectors in
Virasoro
modules in the classical limit[27]. As known, this link is manifest when the
Drinfeld and Sokolovs' matrix representation of differential operators[28] is
exploited. Presumably, a similar link between the superconformally covariant
superdifferential operators and a certain class of singular vectors in
super-Virasoro modules
in the classical limit should also exist. A systematical investigation of this
link would be a very interesting task.
\vskip 1.5cm

\noindent{\bf Acknowledgements}

The author likes to thank Profs. J.C. Shaw and H.C. Yen for valuable
discussions. This work was supported by the National Science Council of
Republic of China under Grant No. NSC-83-0208-M-007-008.

\vskip 1.5cm

\noindent{\bf Note added in proof}

After submitting this work the author became aware of the works by Gieres and
Theisen [29-31]. In ref.[30] the results of sections 4 and 5 of this paper had
been derived in the same spirit and a matrix representation of covariant
superdifferentail operators was also obtained. The author also likes to
recommend refs. [29,30] to those readers who are interested in general aspects
of covariant operators and the relation between classical W-superalgebras and
Lie superalgebras.
\vskip 1.5cm
\noindent{\bf References}

\item{[1]} A.B. Zamolochikov, Theor. Math. Phys. {\bf 65}, 1205 (1985).
\item{[2]} J.-L. Gervais, Phys. Lett. {\bf 160B}, 277 (1985).
\item{[3]} J.-L. Gervais and A. Neveu, Nucl. Phys. {\bf B264}, 557 (1986).
\item{[4]} T.G. Khovanova, Funct. Anal. Appl. {\bf 21}, 332 (1987).
\item{[5]} V.A. Fateev and A.B. Zamolochikov, Nucl. Phys. {\bf B280}, 644
(1987).
\item{[6]} F. A. Bais, P. Bouwknegt, M. Surridge and K. Schoutens, Nucl. Phys.
{\bf 304}, 348, 371 (1988).
\item{[7]} P. Mathieu, Phys. Lett. {\bf 208B}, 101 (1988).
\item{[8]} I. Bakas, Nucl. Phys. {\bf 302}, 189 (1988); Phys. Lett. {\bf 213B},
313 (1988).
\item{[9]} Q. Wang, P.K. Panigraphi, U. Sukhatme and W.-K. Keung, Nucl. Phys.
{\bf B344}, 194 (1990).
\item{[10]} P. DiFrancesco, C. Itzykson and J.-B. Zuber, Comm. Math. Phys.
{\bf 140}, 543 (1991).
\item{[11]} A. Das and S. Roy, Int. J. Mod. Phys. {\bf A6}, 1429 (1991).
\item{[12]} V.A. Fateev and S.L. Lykyanov, Int. J. Mod. Phys. {\bf A3}, 507
(1988).
\item{[13]} A. Das, "Integrable Models', World Scientific, Singapore (1989).
\item{[14]} I.M. Gelfand and L.A. Dickey, Funct. Anal. Appl. {\bf 11}, 93
(1977).
\item{[15]} M. Adler, Invent. Math. {\bf 50}, 219 (1979).
\item{[16]} B.A. Kuperschmidt and G. Wilson, Invent. Math. {\bf 62}, 403
(1981).
\item{[17]} L.A. Dickey, Ann. NY Acad. Sci. {\bf 491}, 131 (1987).
\item{[18]} J.M. Figueroa-O'Farrill, J. Mas and E. Ramos, Phys. Lett. {\bf
266B}, 298 (1991).
\item{[19]} A. Das, W.-J. Huang and S. Panda, Phys. Lett. {\bf 271B}, 109
(1991).
\item{[20]} A.O. Radul in: Applied methods of nonlinear analysis and control,
eds. A. Mironov, V. Moroz and M. Tshernjatin (MGU, Moscow 1987) [in Russian].
\item{[21]} A. Das and W.-J. Huang, J. Math. Phys. {\bf 33}, 2487 (1992).
\item{[22]} J.M. Figueroa-O'Farrill, J. Mas and E. Ramos, A one-parameter
family of hamiltonian structures of the KP hierarchy and a continuous
deformation of the $W_{KP}$ algebra, preprint BONN-HE-92-20, US-FT-92/7,
KUL-TF-922/20 (hep-th/9207092).
\item{[23]} J.M. Figueroa-O'Farrill, J. Mas and E. Ramos, Phys. Lett. {\bf
262B}, 265 (1991).
\item{[24]} J.M. Figueroa-O'Farrill and E. Ramos, Nucl. Phys. {\bf B368}, 361
(1992).
\item{[25]} K. Huitu and D. Nemeschansky, Mod. Phys. Lett. {\bf A6}, 3179
(1991).
\item{[26]} W.-J. Huang, On Diff($S^1$) covariantiztion of pseudodifferential
operator, preprint (hep-th/9309070), to appear in J. Math. Phys.
\item{[27]} M. Bauer, P. DiFrancesco, C. Itzykson and J.-B. Zuber, Nucl. Phys.
{\bf B362}, 515 (1991).
\item{[28]} V.G. Drinfeld and V.V. Sokolov, J. Sov. Math. {\bf 30}, 1975
(1985).
\item{[29]} F. Gieres, Int. J. Mod. Phys. {\bf A8}, 1 (1993).
\item{[30]} F. Gieres and S. Theisen, Superconformally Covariant Operators
and Super W Algebras, preprint MPI-Ph/92-66, KA-THEP-7/92, to appear in
J. Math. Phys.
\item{[31]} F. Gieres and S. Theisen, Classical N=1 and N=2 super W-algebras
from a zero-curvature condition, preprint Mar. 1993, to appear in Int. J.
Mod. Phys. {\bf A}.
\end